\documentclass[12pt]{article}

\newtheorem{thm}{Theorem}
\newtheorem{cor}[thm]{Corollary}

        \def\pf{\medbreak\noindent{\bf Proof:}\enspace}
        
        \def\qed{{\bf QED}}
        \def\iff{\Longleftrightarrow}
        
        \def\tr{\rm {Tr}}

\def\wh{\widehat}
\def\wt{\widetilde}

 \def\ts{\textstyle}
\def\bra{\langle}
\def\ket{\rangle}
\def\kb{ \ket \bra }
\def\dg{\dagger}
\def\ot{\otimes}

\def\raw{\rightarrow}
\def\wh{\widehat}

\def\nl{\newline}
\def\Raw{\Rightarrow}

\def\calm{{\mathcal M}}
\def\cale{{\mathcal E}}

\title{Inequalities for Quantum Entropy: \\
A Review with Conditions for Equality}

\author{Mary Beth Ruskai \\ Department of Mathematics \\
 University of Massachusetts  Lowell \\
Lowell,  MA  01854 USA \\ {\normalsize MaryBeth\_Ruskai@uml.edu} }

\date{\today 
{\normalsize MR classification:  81P68, 94A17, 82B10}}

\begin{document}

\maketitle

\begin{abstract}
This paper presents self-contained proofs of the  strong subadditivity
  inequality for von Neumann's quantum entropy, $S(\rho)$,
and some related inequalities for the quantum relative entropy, most
notably its convexity and its monotonicity under stochastic maps.
Moreover, the approach presented here, which is based on Klein's
inequality and Lieb's theorem  that the function
$A \raw \tr \, e^{K + \log A}$ is concave,  allows one to obtain
conditions for equality.   In the case of strong subadditivity,
which states
  that  $S(\rho_{123}) + S(\rho_2) \leq S(\rho_{12})+S(\rho_{23})$
where the subscripts denote subsystems of a composite system,
equality holds if and only if
$\log \rho_{123} = \log \rho_{12} - \log \rho_2 + \log \rho_{23}$.
Using the fact that the  Holevo bound on the accessible information
in a quantum ensemble can be obtained as a consequence of the
monotonicity of relative entropy, we show
that equality can be attained for that bound only when the
states in the ensemble commute.
The paper concludes with an Appendix giving a short description of
Epstein's elegant proof of   Lieb's theorem.
\end{abstract}



\pagebreak

\tableofcontents


\pagebreak

\section{Introduction}

\subsection{Quantum Entropy}

Quantum information science~\cite{NC} is the study
of the information carrying and processing properties of quantum
mechanical systems.  Recent work in this area has
generated renewed interest in the properties of the quantum mechanical
entropy.
It is interesting to note that von Neumann \cite{vN27,vNbk}
introduced the notion of mixed state, represented by a density
matrix $\rho$ (a positive semi-definite operator with $\tr \rho = 1$),
into quantum theory defined  its entropy  as
 as $ S(\rho) \equiv - \tr(\rho \log \rho)$ in 1927, well before
the corresponding classical quantity was introduced in Shannon's
seminal work \cite{Shan} on ``The Mathematical Theory of Communication''
in 1948.  (Admittedly, von Neumann's motivation was the extension
of the classical theory of statistical mechanics, developed by Gibbs,
Boltzman, et al to the quantum domain rather than the development
of a theory of quantum communication.)  Many fundamental
properties of the quantum entropy were proved in a  remarkable, but
little-known, 1936 paper of Delbr\"uck and Mol\`eiere \cite{DM}.
For further discussion of the history of quantum entropy, see
\cite{OP,Pz.vN,W} and the introductory remarks in \cite{RuSt}.

One important class of inequalities relates
the entropy of subsystems to that of a composite system,
whose Hilbert space is a tensor product
is ${\cal H}_{12} = {\cal H}_1 \ot {\cal H}_2$ of
the Hilbert spaces for the subsystems.    When the
state of the composite system is described by the density
matrix $\rho_{12}$, the states of the subsystems are given
by the reduced density matrices, e.g.,  $\rho_1 = T_2(\rho_{12})$,
obtained by taking the partial trace.   The subadditivity
inequality
\begin{eqnarray} \label{eq:subadd}
S(\rho_{12}) \leq S(\rho_1)+S(\rho_2)
\end{eqnarray}
was proved in \cite{DM} and \cite{LanRob}.  (It should not
be confused with the concavity
\begin{eqnarray} \label{eq:concav}
S(x \rho' + (1-x) \rho'') \geq x S(\rho') + (1-x) S(\rho'')
\end{eqnarray}
which can actually be obtained {\em from}
subadditivity by considering block matrices \cite{LbBull,SSA,W}).
In the more complex situation in which the composite
system is composed of three subsystems the following stronger
inequality, known as  strong subadditivity (SSA), holds.
\begin{eqnarray} \label{eq:ssa}
  S(\rho_{123}) + S(\rho_2) \leq S(\rho_{12}) + S(\rho_{23})
\end{eqnarray}
This inequality was conjectured by
Lanford and Robinson in \cite{LanRob} and proved in \cite{SSA.PRL,SSA}.
In this paper, we review its proof in a form that easily yields
the following condition for equality.
\begin{thm} \label{thm:ssa.eq}
Equality holds in strong subadditivity {\em (\ref{eq:ssa})}
if and only if
 \begin{eqnarray} \label{ssa.equal}
 \log \rho_{123} - \log \rho_{12} = \log \rho_{23} - \log \rho_2.
\end{eqnarray}
\end{thm}
We have suppressed implicit tensor products with the identity
so that,
e.g., $\log \rho_{12}$ means $(\log \rho_{12}) \ot I_3$.
Rewriting (\ref{ssa.equal}) as
$ \log \rho_{123} + \log \rho_2 = \log \rho_{12} + \log \rho_{23}$,
multiplying by $\rho_{123}$ and taking the trace
immediately establishes the sufficiency of this equality condition.
  In Section~\ref{sect:SSA}, we will also show that it is also necessary.

\subsection{Relative entropy} \label{sect:relent}

The SSA inequality can be restated as a property of the
 {\em quantum relative   entropy} which is defined as
\begin{eqnarray} \label{eq:relent}
H(\rho,\gamma) \equiv \tr \, \rho
\big( \log \rho - \log \gamma \big).
\end{eqnarray}
It is usually assumed that $\rho, \gamma$ are density matrices,
although~(\ref{eq:relent}) is well-defined for any pair of positive
semi-definite matrices for which $\ker(\gamma) \subset \ker(\rho)$.
Strong subadditivity can now be restated as
\begin{eqnarray} \label{eq:mono.SSA}
   H(\rho_{12},\rho_{2}) \leq  H(\rho_{123},\rho_{23})
\end{eqnarray}
where we again write, e.g., $\rho_{23}$ for $I_1 \ot \rho_{23}$.
More generally, the relative entropy is monotone
under completely positive,
trace-preserving maps (also known as ``quantum operations''~\cite{NC}
and ``stochastic maps'' \cite{AU,KR1} and discussed in more
detail in section~\ref{sect:Lind.rep}), i.e.,
\begin{eqnarray} \label{eq:mono}
   H[\Phi(\rho),\Phi(\gamma)] \leq  H(\rho,\gamma).
\end{eqnarray}
This monotonicity implies~(\ref{eq:mono.SSA}) when $\Phi = T_3$
is the partial
trace operation; perhaps surprisingly, the converse is also true
\cite{Lind75}.  This, and
other connections between strong subadditivity and relative entropy are
discussed in  Section~\ref{sect:relation}  .

The approach to SSA presented here can also be used to
obtain conditions for equality in  properties of
relative entropy, including its joint convexity and monotonicity.
The explicit statements are postponed to later sections.
Since the monotonicity can be used to give a simple proof of the
celebrated
Holevo bound \cite{Hv0,NC} on accessible information, we show how
our results can be used to recover the equality conditions in
that bound.
As discussed in section~\ref{sect:Petz}, Petz \cite{OP,Pz.eq}
has also obtained several equality conditions in different,
but equivalent, forms.  However, Theorem~\ref{thm:eqal.gen.mono},
which applies to the most general form of monotonicity,
appears to be new.

\subsection{Lieb's convex trace functions}

One of the most frequently cited approaches to strong
subadditivity is to present it as a consequence of the concavity of a
quantity known as the Wigner-Yanase-Dyson entropy \cite{WY}.
This property, conjectured by Bauman \cite{B}, is equivalent
 to the joint concavity
in $A$ and $B$ of the map
\begin{eqnarray}
\label{eq:WYD}
   (A,B) \raw \tr \, A^s K^{\dg} B^{(1-s)} K ~~~ {\rm for} ~~ A, B > 0,
   ~~~ 0 < s < 1
\end{eqnarray}
(where  $\dg$ is used to denote the adjoint).
Lieb's proof~\cite{LbWYD} of the concavity of the WYD function
(\ref{eq:WYD}) and his realization of a connection between SSA and
Bauman's concavity conjecture was a crucial breakthrough.  However,
concavity of the WYD function was only one of several concave trace
functions studied in~\cite{LbWYD}; the
following result was also established by Lieb.
\begin{thm}  \label{exp.conc}
For any fixed self-adjoint matrix $K$, the function
$A \mapsto F(A) = \tr \, e^{K + \log A}$ is concave in $A > 0$.
\end{thm}
This result played a fundamental role in the original
proof~\cite{SSA.PRL,SSA} of SSA and the closely related property
of joint concavity of the relative entropy \cite{SSA.PRL,SSA,Lind74}.
Although SSA is a deep
theorem, a complete proof is not as forbidding as is sometimes implied.
Therefore, for completeness, we include Epstein's elegant proof
\cite{Ep} of
Theorem~\ref{exp.conc} in Appendix~A, and then follow the
original strategies of Lieb and Ruskai~\cite{SSA} to show
how it implies SSA.

\subsection{Overview}

Although this paper grew out of questions about the conditions for
equality in strong subadditivity and related inequalities,
it seems  useful to present these conditions
within a more comprehensive exposition.
For simplicity, we confine our discussion to finite dimensions, and
assume that, unless otherwise stated, the density matrices under
consideration are strictly positive.

The remainder of the paper is structured as follows.
In Section~\ref{sect:reform} we discuss some
consequences and interpretations of the SSA equality condition.
In Section~\ref{sec:tools} we summarize some mathematical
results needed for the proofs in the sections that follow.
Section~\ref{sect:SSA}, which might be regarded as the heart
of the paper, presents the
proof of strong subadditivity in a form which easily yields
 the equality conditions. (A reader primarily interested in
this proof can proceed directly to Section~\ref{sect:SSA} with
a willingness to accept the results of section~\ref{sec:tools}.)
Section~\ref{sect:other} presents proofs with equality conditions for
 the monotonicity of the relative entropy under partial traces, the
joint convexity of the relative entropy; and the general monotonicity
under stochastic maps.  This section also contains a
discussion of the connection between these properties, SSA and their
proofs.   Section~\ref{sect:gen.mono} contains the
proof of the equality conditions for monotonicity of relative entropy.
Section~\ref{sect:Holv} consider bounds, most notably the Holevo bound, on
the accessible information that can be extracted from an ensemble
of quantum states, and the conditions under which they can
be attained. The
paper concludes with some additional historical
comments in Section~\ref{sect:conc}.

\section{Implications of the equality conditions for SSA}
     \label{sect:reform}

\subsection{Classical conditions} \label{sect:class}

To describe the corresponding classical inequalities,
let the subsystems $A,B$ and $C$ correspond to classical
random variables.  One can recover
the classical Shannon entropy $ - \sum_a p(a) \log p(a) $ from the von
Neumann entropy by taking $\rho$ to be a diagonal matrix with elements
$p(a)$ on the diagonal.  Employing a slight abuse of notation, we
write $S[p(a)]$ for this quantity. Then the classical strong
subadditivity inequality can be stated as
\begin{eqnarray} \label{eq:SSA.class}
 S[p(a,b,c)] +  S[p(b)] \leq S[p(a,b)] + S[p(b,c)] .
\end{eqnarray}
The classical relative entropy of the distribution $q(a)$
with respect to $p(a)$ is
$H[p(a),q(a)] = \sum_a p(a) \log \frac{p(a)}{q(a)}$.
It is well-known  (see, e.g.,\cite{Kull}) that
the convexity of the function $f(x) = x \log x$ implies
that $H[p(a),q(a)] \geq 0$ and its strict convexity
implies that equality holds if and only if $p(a) = q(a) ~~ \forall~a$.
(The generalization of this result to quantum situations is
discussed in section \ref{sect:klein}.)

The classical form (\ref{eq:SSA.class}) of SSA is equivalent to
$H[p(a,b,c), q(a,b,c)] \geq 0$ when the second
distribution is $q(a,b,c) = p(a,b) [p(b)]^{-1} p(b,c)$,
Thus, equality holds in (\ref{eq:SSA.class})
if and only if
\begin{eqnarray} \label{eq:SSA.class.prod}
 p(a,b,c) = p(a,b) [p(b)]^{-1} p(b,c)   ~~~~ \forall~a,b,c
\end{eqnarray}
which can be rewritten as
\begin{eqnarray}  \label{eq:SSA.class.eqal}
\log p(a,b,c) - \log p(a,b) =  \log p(b,c) - \log p(b)
 ~~~~ \forall~a,b,c.
\end{eqnarray}
which is identical to what one would obtain
from Theorem~\ref{thm:ssa.eq}.  Using $p(c|b)$ to denote
 the classical conditional probability distribution,
(\ref{eq:SSA.class.eqal}) can be rewritten  as
\begin{eqnarray} \label{eq:classical_ssa}
p(c|a,b) = p(c|b),
\end{eqnarray}
which is precisely the condition
that the sequence $A \rightarrow B \rightarrow C$ forms a Markov
chain.

\subsection{Special cases of SSA equality} \label{sect:equal}

Some insight into equality condition~(\ref{ssa.equal}) may be
obtained by looking at
special cases in which it is satisfied.  The most obvious
 is when $\rho_{123}$ is a tensor product of its three
reduced density matrices.  However,  it is readily verified that
(\ref{ssa.equal}) also holds  when either $\rho_{123} = \rho_1 \ot
\rho_{23}$ {\em or} $\rho_{123} = \rho_{12} \ot \rho_{3}$.  One
can generalize this slightly further.  If the subsystem $2$
 can be  partitioned further into two subsystems $2'$
and $2''$, then one can verify  equality holds if
$\rho_{123} = \rho_{12'} \otimes \rho_{2''3}$, where $\rho_{12'}$
and  $\rho_{2''3}$ are
states of the composite systems $1,2'$ and
 $2'',3$ respectively.

However,  such a
decomposition into tensor products is not necessary; indeed, we
have already seen that
 equality also holds for the case of classical Markov
processes.  Moreover, by comparison to (\ref{eq:classical_ssa})
it is natural to regard
(\ref{ssa.equal}) as a kind of quantum Markov
condition.
Thus, the  conditions in Theorem~\ref{thm:ssa.eq} can also be viewed as a
natural non-commutative analogue of the conditions for equality in
classical SSA.
Another way of regarding~(\ref{ssa.equal}) is as a concise statement
of a subtle intertwining condition discussed below.
Unfortunately, we have not found explicit
 examples which satisfy it other than the two classes discussed
above, that is, a partial decomposition into tensor products or a
classical Markov chain.

\subsection{Petz's conditions} \label{sect:Petz}

Using a completely different approach,
Petz~\cite{OP,Pz.eq} gave conditions for equality in
(\ref{eq:mono}) when $\Phi$ can be identified with a mapping
 of an algebra onto a subalgebra,
a situation which includes (\ref{eq:mono.SSA}).
In that case Petz's conditions become
\begin{eqnarray} \label{eq:mono.Pz.eqal}
   \rho_{12}^{it} \rho_2^{-it}  =  \rho_{123}^{it} \rho_{23}^{-it}.
\end{eqnarray}
Taking the derivative of both sides of (\ref{eq:mono.Pz.eqal})
at $t = 0$ yields (\ref{ssa.equal}).
Although (\ref{eq:mono.Pz.eqal}) appears stronger than (\ref{ssa.equal}),
it is not since, as noted above,
(\ref{ssa.equal}) is sufficient for equality in (\ref{eq:mono.SSA}).
Moreover, since  (\ref{ssa.equal}) implies
\begin{eqnarray}
   e^{i t \, \log ( \rho_{123})}  & = &
  e^{i t \, \left[ \log \rho_{12} - \log \rho_2 + \log \rho_{23}\right]}
\end{eqnarray}
our results can be combined with those of Petz to see that equality
holds in SSA $\iff$ (\ref{ssa.equal}) $\iff$ (\ref{eq:mono.Pz.eqal})
and that any of these conditions suffices to imply
\begin{eqnarray}
e^{i t \, \left[ \log \rho_{12} - \log \rho_2 + \log \rho_{23}\right]}
=  e^{i t \, \log ( \rho_{12})} e^{-i t \, \log ( \rho_{2})}
e^{i t \, \log ( \rho_{23})} .
\end{eqnarray}
Note that one can also relate Petz's conditions
to  those for equality in classical  SSA
by rewriting (\ref{eq:SSA.class.prod}) as
$ p(a,b,c)[p(b,c)]^{-1} = p(a,b) [p(b)]^{-1}$ and
then raising
to the $it$ power.

\section{Fundamental mathematical tools}  \label{sec:tools}

\subsection{Klein's inequality} \label{sect:klein}

The fact that the relative entropy is positive, i.e.,
$H(\rho,\gamma) \geq 0$ when $\tr \, \rho = \tr \, \gamma$ is an
immediate consequence of the following fundamental
convexity result due to
Klein~\cite{Kl,NC,W}.
\begin{thm}
\label{thm:klein} {\em (Klein's Inequality)} For $A, B > 0$
\begin{eqnarray} \label{ineq:klein}
   \tr \, A \big( \log A -  \log B \big) \geq \tr (A - B),
\end{eqnarray}
with equality if and only if $A = B$.
\end{thm}
The closely related Peierls-Bogoliubov  inequality \cite{OP,W}
is sometimes
 used instead of Klein's inequality.  However, the equality
conditions in Theorem~\ref{thm:klein}  play a critical role
in the sections that follow.

\subsection{Lieb's golden corollary}

The proofs in Section~\ref{sect:SSA} do not use Theorem~\ref{exp.conc}
directly,  but a related result  generalizing the
following inequality, which we will also need.
\begin{thm} \label{thm:gold} {\em (Golden-Thompson-Symanzik)}
For self-adjoint matrices $A$ and $B$
$\tr \, e^{A+B} \leq \tr \, e^A e^B$ with equality if and
only if $A$ and $B$ commute.
\end{thm}
Although this inequality is extremely well-known,
the conditions for equality do not appear explicitly in such
standard references as \cite{HJ2,Sim,W}.   However,
one method of proof is based on the observation that
$ \tr \, [e^{A/2^k} e^{B/2^k}]^{2^k}$ is monotone
decreasing in $k$, yielding $e^{A+B}$ in the limit as
$k \raw \infty$.  The equality conditions then follow easily
from those for the Schwarz inequality for  the
 Hilbert-Schmidt inner product $\tr C^{\dg} D$.
Indeed,  $k = 1$ yields
\begin{eqnarray*}
   \tr \Big( e^{A/2} e^{B/2} \Big) \Big( e^{A/2} e^{B/2} \Big)  \leq
  \Big[ \tr \, e^{B/2} e^A  e^{B/2} \Big]^{1/2}
    \Big[ \tr \, e^{A/2} e^B  e^{A/2} \Big]^{1/2} = \tr \, e^A e^B
\end{eqnarray*}
 with $C = e^{B/2} e^{A/2}$ and $D = e^{A/2} e^{B/2}$.
The equality  condition  that $C$ is a multiple of $D$ implies
$e^{B/2} e^{A/2} = e^{A/2} e^{B/2}$ which holds if and
only if  $A$ and $B$ commute.  One  reference \cite{OP} that
does discuss equality does so by making the
interesting observation that (as shown in \cite{Pz.gt})
Theorem~\ref{thm:gold} and its equality conditions, can be
derived as a consequence of the monotonicity of relative entropy,
Theorem~\ref{eq:mono}.

The natural extension  to three matrices
 $\tr \, e^{A + B + C} \leq |\tr e^A e^B e^C|$,
fails; see, for example, Problem~20 on pages 512--513
of~\cite{HJ2}.   Therefore, the following result
of Lieb~\cite{LbWYD} is particularly noteworthy.
\begin{thm} {\em (Lieb)} \label{cor:trip.gold}
For any $R, S, T > 0$ \begin{eqnarray}
\label{eq:trip.gold}
  \tr \, e^{\log R - \log S + \log T} \leq
  \tr \, \int_0^\infty R \frac{1}{S + u I} T  \frac{1}{S + u I} du.
\end{eqnarray}
\end{thm}
One might expect that equality holds if and only if $R,S, T$
commute.  Although this is sufficient, it is not necessary.
One easily checks that both sides of (\ref{eq:trip.gold})
equal $\tr \, \rho_1 \ot \rho_{23}$ when
$R = \rho_1 \ot \rho_2 \ot I_3, S = I_1 \ot \rho_2 \ot I_3,
 T = I_1 \ot \rho_{23}$, even when $T$ does not commute with $R$
or $S$.
 \pf Lieb's proof of (\ref{eq:trip.gold}) begins with the
easily-established fact \cite{Rock} that if
$F(A)$ is concave and homogeneous in the sense $F(xA) = x F(A)$ , then
\begin{eqnarray} \label{conc.deriv.ineq}
   \lim_{x \raw 0} \frac{F(A + xB) - F(A)}{x}  \geq F(B) .
\end{eqnarray}
Applying this to the functions in Theorem~\ref{exp.conc} with
$A = S, B = T, K = \log R - \log S$ yields
\begin{eqnarray} \label{eq:intermediate_Lieb}
\tr e^{\log R-\log S+\log T} \leq \lim_{x\rightarrow 0}
\frac{\tr e^{\log R-\log S+\log(S+xT)}-\tr R}{x}.
\end{eqnarray}
To complete the proof, we need the well-known integral representation
\begin{eqnarray} \label{eq:int_repn}
  \log(S + xT) - \log S = \int_0^\infty
   \frac{1}{S + u I} \, xT  \, \frac{1}{S + xT + u I} \,  du ~.
\end{eqnarray}
Substituting (\ref{eq:int_repn}) into (\ref{eq:intermediate_Lieb})
and noting that
\begin{eqnarray*}
\tr \, e^{\log R + x \int_0^\infty
   \frac{1}{S + u I} \, T  \, \frac{1}{S + xT + u I} \,  du} =
 \tr \, R + x \, \tr \, R \int_0^\infty \!
   \frac{1}{S + u I} \, T  \, \frac{1}{S + u I} \,  du + O(x^2)
\end{eqnarray*}
yields the desired result.
~~~ \qed

\subsection{Purification}  \label{sect:pure}
Araki and Lieb \cite{AkLb,LbBull}
observed that one could obtain useful new entropy inequalities
by applying what is now known as the
``purification process'' to known inequalities.
Any density $\rho_1$ can be extended to a pure state
density matrix $\rho_{12}$ on a tensor product space;
moreover, $S(\rho_1) = S(\rho_2)$.  Applying
this to the subadditivity inequality (\ref{eq:subadd}), i.e.,
$S(\rho_{12}) \leq S(\rho_1)+S(\rho_2)$,
yields the equivalent result $S(\rho_3) \leq  S(\rho_{23})+S(\rho_2)$
which can be combined with (\ref{eq:subadd}) to give the triangle
inequality \cite{AkLb,LbBull}
\begin{eqnarray}  \label{eq:tri}
| S(\rho_1) - S(\rho_2)| \leq S(\rho_{12}) \leq S(\rho_1)+S(\rho_2).
\end{eqnarray}
By purifying $\rho_{123}$ to $\rho_{1234}$ one can similarly
show that SSA (\ref{eq:ssa}) is equivalent to
\begin{eqnarray}  \label{eq:ssa.alt}
   S(\rho_4) +  S(\rho_2) \leq S(\rho_{12}) +  S(\rho_{14}).
\end{eqnarray}


\subsection{Lindblad's representation of stochastic maps}
 \label{sect:Lind.rep}

Stochastic
maps arise naturally in quantum information as a description
of the effect on a subsystem $A$
interacting with the environment in the pure state
$\gamma_B = |\psi_B \kb \psi_B|$ via the unitary operation $U_{AB}$,
\begin{eqnarray}
   \rho_A \raw
\tr_B \Big( U_{AB} \, \rho_A \otimes \gamma_B \, U_{AB}^{\dagger}  \Big).
\end{eqnarray}
Lindblad~\cite{Lind75} used Stinespring's representation
to show that any completely positive
trace-preserving map
$\Phi$ which maps an algebra into itself can be represented as if it
arose in this way.  That is, given such a map
$\Phi$ one can always find an auxiliary system, ${\cal H}_B$, a
density matrix
$\gamma_B$ on ${\cal H}_B$, and a unitary map $U_{AB}$ on the
combined system ${\cal H}_A \ot {\cal H}_B$
(where $A$ denotes the original system) such
that
\begin{eqnarray} \label{eq:Lind.rep}
     \Phi(\rho) = \tr_B
       \Big( U_{AB} \, \rho \ot \gamma_B U_{AB}^{\dagger} \Big)
\end{eqnarray}
where $\tr_B$ denotes the partial trace over the auxiliary system.

Using the Kraus representation $\Phi(\rho) = \sum_k F_k \rho
F_k^{\dg}$  (and noting that the
requirement that $\Phi$ be trace-preserving is equivalent to
 $\sum_k F_k^{\dg}F_k = I$), one can give a construction
equivalent to Lindblad's by initially
defining  $U_{AB}$ as
\begin{eqnarray} \label{eq:unit.def}
U_{AB}|\psi \ket  \ot |\beta \ket \equiv \sum_k F_k|\psi \ket \ot
|k \ket,
\end{eqnarray}
where $|\beta \ket$ is a fixed normalized state of the auxiliary
system, and $\{ |k \ket \}$ is some orthonormal basis for the
auxiliary system.  Then $U_{AB}$ is a partial isometry from
${\cal H}_A \ot |\beta \kb \beta|$ to  ${\cal H}_A \ot {\cal H}_B$
which can be extended to a unitary operator
 on all of ${\cal H}_A \ot {\cal H}_B$.
This yields (\ref{eq:Lind.rep}) with $\gamma_B = |\beta \kb \beta|$
a pure state.

However, $U_{AB}$ can also be extended to  ${\cal H}_A \ot {\cal H}_B$
in other ways.   In particular, it can be extended, instead, to the
partial isometry for which $U_{AB}^{\dg} U_{AB}$
is the projection onto ${\cal H}_A \ot |\beta \kb \beta|$
so that $U_{AB} = 0$ on the orthogonal complement of
${\cal H}_A \ot |\beta \kb \beta|$.  We describe this in
more detail when $\Phi$ requires at most $m$ Kraus
operators $F_k$, in which case one can
 choose the auxiliary  system to be ${\bf C}^m$.
One can also choose  $| k \ket = |e_k \ket$,
 and $| \beta \ket = | e_1 \ket$ with $ |e_k \ket$ the
standard basis of column vectors
with elements $c_j = \delta_{jk}$.   Then
 (\ref{eq:unit.def}) depends only on the first column of
$U_{AB}$ which we denote $V$ and regard as a   map from ${\cal H}$
to ${\cal H} \ot {\bf C}^m$.  In block form
\begin{eqnarray}   \label{eq:Vrep}
V \rho V^{\dg}  & = & U_{AB} \, \rho \ot |e_1 \kb e_1 | U_{AB}^{\dagger}
  \\ & ~ & \nonumber \\ & = &
\pmatrix{ F_1 \cr F_2 \cr ~ \cr \vdots \cr ~ \cr F_m} \rho
\pmatrix{ F_1^{\dg} & F_2^{\dg} & \ldots &  F_m^{\dg} }
 = \pmatrix{ F_1 \rho F_1^{\dg} & F_1 \rho F_2^{\dg} & \ldots &
 F_1 \rho F_m^{\dg} \cr F_2 \rho F_1^{\dg}  & F_2 \rho F_2^{\dg}
   &  \ldots & F_2 \rho F_m^{\dg} \cr  ~ & ~ & ~ & ~ \cr
   \vdots & \vdots & ~ & \vdots \cr   ~ & ~ & ~ & ~ \cr
 F_m \rho F_1^{\dg} & ~ & \ldots & F_m \rho F_m^{\dg}} \nonumber
\end{eqnarray}
from which it easily follows that
$\tr_B (V \rho V^{\dg}) = \sum_k F_k \rho F_k^{\dg} = \Phi(\rho)$.
The requirement that $\Phi$ be trace-preserving gives
$V^{\dg} V = \sum_k F_k^{\dg} F_k = I$ which again implies
that $V$ is a partial isometry.
Moreover, $V \rho V^{\dg}$
 has the same non-zero eigenvalues as
$ (V \sqrt{\rho})^{\dg} (V \sqrt{\rho}) = \rho $ so that
$S[V \rho V^{\dg}] = S(\rho)$.

This construction can be readily extended to situations in
which $\Phi$ maps operators acting on one Hilbert space
${\cal H}_A$ to those acting on another space ${\cal H}_{A'}$,
e.g., $\Phi : {\cal B}({\cal H}_A) \mapsto {\cal B}({\cal H}_{A'})$.
In this case, the Kraus operators
$F_k : {\cal H}_A \mapsto {\cal H}_{A'}$, and $U_{AB}$ is a
partial isometry from ${\cal H}_A \ot |\beta \kb \beta|$ to a
subspace of $ {\cal H}_{A'} \ot {\cal H}_B$.  Alternatively,
$V$ can be defined as a partial isometry from ${\cal H}_A$ to
$ {\cal H}_{A'} \ot {\bf C}^m$.

\subsection{Measurements and their representations}
 \label{sect:POVM}
A von Neumann or {\em projective measurement} is a partition
of the identity $I = \sum_b E_b$ into mutually orthogonal
projections, i.e., $E_b E_c = \delta_{bc} E_b$.
A positive operator valued measurement (POVM)  is a set
of positive semi-definite operators $E_b$ such that
$\sum_b E_b = I$, i.e., the orthogonality condition is dropped.
It is well-known that a general POVM can be represented
as a projective measurement on a tensor product space \cite{NC}.

In fact, by  noting that the map
$\rho \mapsto \sum_b \sqrt{E_b} \,  \rho \sqrt{E_b}$ is completely
positive and trace-preserving with Kraus operators $F_b = \sqrt{E_b}$
one use the construction above.
Write $V = \sum_b  \sqrt{E_b} \ot | b \ket$ where $| b \ket$
is an orthonormal basis for ${\bf C}^M$ and $M$ is the number of
measurements in the POVM, i.e., $b = 1 \ldots M$.   Then
$V \rho V^{\dg} =
   \sum_{b,c} \sqrt{E_b} \, \rho \, \sqrt{E_c} \ot | b \kb c|$.
 Now, if $F_b = I \ot |b \kb b|$,
then $\{ F_b \}$ is a projective measurement on
${\cal H} \ot {\bf C}^M$ and
$\tr\,  F_b \, (V \rho V^{\dg})  = \tr \, E_b \rho$.

\subsection{Adjoint maps} \label{sect:adj}

It is sometimes useful to consider the
adjoint, which we denote $\wh{\Phi}$, of a stochastic
map $\Phi$ with respect
to the Hilbert-Schmidt inner product $\bra A, B \ket = \tr \, A^{\dg} B$.
When $\Phi$ acts on $n \times n$ matrices, this adjoint (or dual)
is fully defined by the requirement
\begin{eqnarray}  \label{eq:HSadj}
  \tr \, [\Phi(A)]^{\dg} B = \tr \, A^{\dg} \wh{\Phi}(B).
\end{eqnarray}
for all $n \times n$ matrices, $A,B$.  Indeed, when
$\Phi(\rho) = \sum_k F_k \rho F_k^{\dg}$, the adjoint
is given by  $\wh{\Phi}(\rho) = \sum_k F_k^{\dg} \rho F_k$.
Moreover, $\Phi$ is trace-preserving if and only if 
$\wh{\Phi}$ is unital, i.e, $\wh{\Phi}(I) = I$.
 When $\Phi$ is the
partial trace, $T_2$, its adjoint takes $A \mapsto A \ot I_2$.


\section{Subadditivity proofs} \label{sect:SSA}

To understand the proof of strong subadditivity, it is instructive to
first understand how Klein's inequality can be used to prove two
weaker inequalities.  First, we consider the subadditivity inequality
(\ref{eq:subadd}).
Substituting $A = \rho_{12}$ and $B = \rho_1 \ot \rho_2$
into Klein's inequality~(\ref{ineq:klein}) yields
\begin{eqnarray}
   - S( \rho_{12} ) + S(\rho_1) + S(\rho_2) \geq
   \tr \big( \rho_{12} - \rho_1 \ot \rho_2 \big) = 0,  \end{eqnarray}
which is equivalent to subadditivity.
Furthermore, the well-known conditions for equality in subadditivity
follow from the conditions for equality in Klein's inequality, namely
that equality holds if and only if $\rho_{12}$ is a tensor product, that is,
$\rho_{12} = \rho_1 \ot \rho_2$.

%
%

A second, more powerful subadditivity inequality was obtained by Araki
and Lieb~\cite{AkLb},
\begin{eqnarray} \label{eq:Arak.Lb}
S( \rho_{123} ) \leq S(\rho_{12} ) + S(\rho_{23} )
\end{eqnarray}
under the constraint  $\tr \rho_{123} = 1$.  To prove this,
choose $A = \rho_{123}$ and $B = e^{\log \rho_{12} + \log \rho_{23}}$
in Klein's inequality to obtain
\begin{eqnarray}
  - S( \rho_{123} ) + S(\rho_{12} ) + S(\rho_{23} ) & \geq &  1 -
   \tr e^{\log \rho_{12} + \log \rho_{23}}.
\end{eqnarray}
Applying Theorem~\ref{thm:gold}, to the right-hand side gives
\begin{eqnarray*}
  - S( \rho_{123} ) + S(\rho_{12} ) + S(\rho_{23} ) & \geq &
   1 -  \tr_{123} \rho_{12} \rho_{23} \\
  & = & 1 - \tr_2 (\rho_2)^2 \\
  & \geq & 1-\tr_2 \rho_2 = 0,
\end{eqnarray*}
where the last line follows from $(\rho_2)^2 \leq \rho_2$
(which is the {\em only}
place  the normalization condition $\tr \rho_{123} = 1$
is needed).   \qed

The strategy for proving SSA is similar to that above, but
with Theorem~\ref{thm:gold} replaced by Theorem~\ref{cor:trip.gold}.
Let $A = \rho_{123}$ and choose
$B$ so that $\log B = \log \rho_{12} - \log \rho_2 + \log \rho_{23}$.
Then Klein's inequality implies
\begin{eqnarray}
\lefteqn{  - S( \rho_{123} ) + S(\rho_{12} ) - S(\rho_2) + S(\rho_{23} ) }
   ~~~~~~~~~~~~~~~~~~~~~~ \nonumber \\
& \geq &
  \tr \left( \rho_{123} - e^{ \log \rho_{12} - \log \rho_2 + \log
\rho_{23}} \right).
\end{eqnarray}
Applying Lieb's result~(\ref{eq:trip.gold}) to the right-hand side above,
we obtain
\begin{eqnarray*}
\lefteqn{  - S( \rho_{123} ) + S(\rho_{12} ) - S(\rho_2) + S(\rho_{23} ) }
   ~~~~~~~~~~~~~~~~~~~~~~ \nonumber \\
  & \geq & \tr \left( \rho_{123} -  \int_0^\infty \rho_{12} \frac{1}{\rho_2
+ u I}
 \rho_{23} \frac{1}{\rho_2 + u I} du \right) \\ & = &
 \tr_{123} \, \rho_{123} -
    \tr_2 \int_0^\infty \rho_{2} \frac{1}{\rho_2 + u I}
  \rho_{2} \frac{1}{\rho_2 + u I} du\\
   & = &  \left( \tr_{123} \, \rho_{123} - \tr_2  \rho_2 \right)  = 0.
\end{eqnarray*}
This proves SSA.  Moreover, this approach allows us to easily
determine the conditions for equality, and thus complete the proof of
Theorem~\ref{thm:ssa.eq}.  The first inequality in the derivation
above is satisfied with equality if and only if $A = B$ which is just
the condition~(\ref{ssa.equal}).  Although the conditions for equality
in~(\ref{eq:trip.gold}) are more difficult to analyze, this is not
necessary here.  When $A = B$, it immediately follows that $\tr \, A =
\tr \, B$ so that the second inequality in the above derivation
automatically becomes an equality when~(\ref{ssa.equal}) holds.

\section{Inequalities for relative entropy } \label{sect:other}

\subsection{Monotonicity under partial trace}
\label{sect:mono}

We now show how the same strategy can be applied to obtain a proof with
equality conditions for the monotonicity of relative entropy under
partial trace.
\begin{thm} \label{thm:mon.tr}
When $\rho_{12},\gamma_{12} > 0$ and $\tr \rho_{12} = \tr \gamma_{12}$
\begin{eqnarray} \label{eq:mon.tr}
 H(\rho_2,\gamma_2) \leq H(\rho_{12},\gamma_{12})
\end{eqnarray}
with equality if and only if
$ \log \rho_{12} - \log \gamma_{12} = \log \gamma_2 + \log \rho_{2} $.
\end{thm}
This condition  should be interpreted as
$ \log \rho_{12} - \log \gamma_{12} = I_1 \ot
   \Big[ \log \gamma_2 - \log \rho_{2} \Big] $.
Since, as noted in section~\ref{sect:adj}, when $\Phi = T_1$, the action
$\wh{\Phi}$ is  precisely  $I_1 \ot $, the equality condition can be
written as
$  \log \rho_{12} - \log \gamma_{12} = \wh{T}_1
   \Big[ \log T_1(\gamma_{12}) - \log T_1(\rho_{12}) \Big]$
which  is a special case of  the more general
form  (\ref{eq:eqal.gen.mono}) developed later.

SSA can be regarded as a special case of
this monotonicity result via the correspondence
$\rho_{12} \raw \rho_{123}, ~\gamma_{12} \raw \rho_{12}$, and
Petz's form of the equality condition becomes
$  \rho_{2}^{it} \gamma_2^{-it}  =  \rho_{12}^{it}  \gamma_{12}^{-it}$.
It is interesting to note that in \cite{SSA}, Lieb and Ruskai
actually obtained
equation~(\ref{eq:mon.tr}) {\em from} SSA using the convexity of
the conditional entropy $S(\rho_1) - S(\rho_{12})$ and the
inequality~(\ref{conc.deriv.ineq}).

\noindent{\bf Proof:} Let
$A = \rho_{12}$,
$\log B = \log \gamma_{12} - \log \gamma_2 + \log \rho_{2}$.
Then Klein's inequality and~(\ref{eq:trip.gold}) imply
\begin{eqnarray*}
  H(\rho_{12},\gamma_{12}) - H(\rho_2,\gamma_2) & \geq & \tr_{12} \left(
\rho_{12}-
        e^{\log \gamma_{12} - \log \gamma_2 + \log \rho_{2}} \right)  \\
& \geq & \tr_{12}  \left( \rho_{12} -  \int_0^\infty \gamma_{12}
\frac{1}{\gamma_2 + uI}  \rho_{2} \frac{1}{\gamma_2 + u I} du \right) \\ & = &
 \tr_{12} \,  \rho_{12} -  \tr_{ 2}\int_0^\infty \gamma_{2}
\frac{1}{\gamma_2 + u I}
 \rho_{2} \frac{1}{\gamma_2 + u I} du  \\ & = &
\tr_{12}  \, \rho_{12}  - \tr_2   \rho_{2}  = 0.
  \end{eqnarray*}
The equality condition is again precisely the condition $A =B$. ~~\qed

\subsection{Joint convexity of the relative entropy}
\label{sect:convexity}

The joint
convexity of relative entropy can  be obtained directly from
Theorem~\ref{thm:mon.tr} by choosing $\rho_{12}$ (and similarly
$\gamma_{12}$) to be a block diagonal matrix with blocks $\lambda_k
\rho^{(k)}$ (and $\lambda_k \gamma^{(k)}$).  We can interpret the
partial trace as a sum over blocks so that $\rho \equiv \rho_2 =
\sum_k \lambda_k \rho^{(k)}$.  However, it is worth giving a direct
proof of the joint convexity since it demonstrates the central role of
Theorem~\ref{exp.conc}.
\begin{thm} \label{thm:relent.conv}
The relative entropy is jointly convex in its arguments, i.e., if
$\rho = \sum_k \lambda_k \rho^{(k)}$ and
$\gamma = \sum_k \lambda_k \gamma^{(k)}$, then
\begin{eqnarray} \label{eq:relent.conv}
H(\rho, \gamma)  \leq
\sum_k \lambda_k \,  H \left( \rho^{(k)} \,  , \, \gamma^{(k)} \right)
\end{eqnarray}
with equality if and only if
$\log \rho - \log \gamma = \log  \rho^{(k)} - \log \gamma^{(k)}$
for all $k$.
\end{thm}

\noindent{\bf Proof:}  Let $A =  \rho^{(k)}$
and $\log  B = \log  \rho -  \log \gamma+ \log  \gamma^{(k)} $
with $\rho = \sum_k \lambda_k \rho^{(k)}$ and
$\gamma = \sum_k \lambda_k \gamma^{(k)}$. Then Klein's inequality implies
\begin{eqnarray}
  H \left( \rho^{(k)} \,  , \, \gamma^{(k)} \right) -
    \tr \, \rho^{(k)} \big[ \log \rho  - \log \gamma \big] \geq
 \tr \left( \rho - e^{\log \rho - \log  \gamma + \log  \gamma^{(k)}}
      \right)
\end{eqnarray}
Multiplying this by $\lambda_k$ with $\lambda_k > 0 $ and
$\sum_k \lambda_k = 1$ yields, after summation,
\begin{eqnarray*}
\lefteqn{  \sum_k \lambda_k \,  H \left( \rho^{(k)} \,  , \, \gamma^{(k)}
\right)
   - H(\rho,\gamma) } ~~~~~~~~~~~~   \\
  & \geq &      \tr \left( \rho - \sum_k \lambda_k
  e^{\log \rho - \log  \gamma  + \log  \gamma^{(k)}} \right)  \\
    & \geq & \tr \left( \rho -
  e^{\log \rho - \log  \gamma  + \log  \sum_k \lambda_k \gamma^{(k)}}
\right)  \\
  & = & \tr \left( \rho - e^{\log \rho} \right) = 0
\end{eqnarray*}
where the second inequality is precisely the concavity of
$C \raw F(C) = \tr e^{K + \log C}$ with $K = \log \rho - \log  \gamma$ and
$ C =  \sum_k  \lambda_k \gamma^{(k)}$.

\subsection{Relationships among inequalities} \label{sect:relation}

We make some additional remarks about connections between
 SSA and various properties of relative entropy.  To facilitate the
discussion, we will use MONO to denote the general monotonicity inequality
(\ref{eq:mono}), MPT to denote the special case of monotonicity under
partial traces, i.e., Theorem~\ref{thm:mon.tr}, and JC to denote the
joint convexity, Theorem~\ref{thm:relent.conv}.  Using the restatement
of SSA in the form (\ref{eq:mono.SSA}), it is easy to see that MONO
$\Raw$ MPT $\Raw$ SSA.  Before theorem~\ref{thm:relent.conv}, we
showed that MPT $\Raw$ JC.
Similarly, by choosing $\rho_{123}$ to be
block diagonal with blocks $\rho_{123}^{k}$ one can show that
SSA implies that the map $\rho_{12} \mapsto S(\rho_1) - S(\rho_{12} )$
is convex.  In \cite{SSA} it was observed that applying the convexity
inequality  (\ref{conc.deriv.ineq}) to this map
(with $A + xB = \rho_{12} + x \gamma_{12}$),
yields (\ref{eq:mon.tr}).  This shows that SSA $\Raw$ MPT so that
we have the chain of implications
\begin{eqnarray} \label{eq:chain}
     \hbox{MONO} \Raw \hbox{MPT}  \iff SSA \Raw \hbox{JC}.
\end{eqnarray}
One can show that JC $\Raw$ MPT by using Uhlmann's observation
\cite{Uhl1} that the partial trace can be written as a convex
combination of unitary transformations.

One can also show directly that JC $\Raw$ SSA by  using the purification
process described in section~\ref{sect:pure}
to show that SSA is equivalent to
\begin{eqnarray} \label{eq:SSA.alt}
     \rho_4 + \rho_2 \leq  \rho_{12} + \rho_{14} .
\end{eqnarray}
Moreover, if $\rho_{124}$ is pure, then $\rho_4 =  \rho_{12}$ and
$\rho_2 =  \rho_{14}$ so that equality holds in (\ref{eq:SSA.alt}).
Since the extreme points of the convex set of density matrices
are pure states, the inequality (\ref{eq:SSA.alt}) then follows
from the joint convexity, Theorem~\ref{thm:relent.conv}.  Thus we
have
\begin{eqnarray} \label{eq:chain2}
     \hbox{MONO} \Raw \hbox{MPT}  \iff SSA \iff \hbox{JC}.
\end{eqnarray}
Lindblad~\cite{Lind75} completed this circuit by showing that
MPT $\Raw$ MONO.

Using  the representation described in
 Section~\ref{sect:Lind.rep},
 with $V$ the partial isometry from ${\cal H}$ to ${\cal H} \ot {\bf
C}^m$  as in (\ref{eq:Vrep}), one finds
\begin{eqnarray}
  H\big[\Phi(\rho), \Phi(\gamma) \big] & = &
H \Big[ \, \tr_B \big( V  \rho  V^{\dg} \big) \, ,
    \, \tr_B \big( V  \gamma V^{\dg} \big)  \, \Big]
 \nonumber \\ \label{eq:mono.via.pt} & \leq &
  H \Big[\,V  \rho \, V^{\dg}  \, ,  V \gamma \,V^{\dg} \, \Big] \\
    & = &   H(\rho,\gamma)
\end{eqnarray}
since
$\tr V \rho V^{\dg} \log (V \gamma V^{\dg}) =  \tr \, \rho \log \gamma$
for  a partial isometry $V$.

\section{Equality in monotonicity under stochastic maps} \label{sect:gen.mono}

Conditions for equality in the general monotonicity
inequality (\ref{eq:mono}) may be more subtle since it is not
always possible to achieve equality.  Indeed, it was noted
in ~\cite{LesRu} that
$  \sup_{\rho \neq \gamma} \frac{H[\Phi(\rho) ,\Phi(\gamma)] }
{H(\rho,\gamma) } $ can be strictly less than 1.  Using
the reformulation (\ref{eq:mono.via.pt}) above, we prove
the following result.
\begin{thm}  \label{thm:eqal.gen.mono}
Equality holds in {\em (\ref{eq:mono})},
$H\big[\Phi(\rho), \Phi(\gamma) \big] \leq H(\rho,\gamma)$,
if and only if
\begin{eqnarray}  \label{eq:eqal.gen.mono}
  \log \rho - \log \gamma = \wh{\Phi}
   \left[  \log \Phi(\rho) - \log \Phi(\gamma) \right] \label{eq:cond.d}
\end{eqnarray}
where $\widehat{\Phi}$ denotes the adjoint of $\Phi$ with respect
to the Hilbert-Schmidt inner product as defined in (\ref{eq:HSadj}).
\end{thm}
To verify sufficiency, multiply (\ref{eq:eqal.gen.mono}) by
$\rho$ and take the trace to obtain
\begin{eqnarray*}
  H[\rho,\gamma] & = &  \tr \, \rho \, \wh{\Phi}
   \left[  \log \Phi(\rho) - \log \Phi(\gamma) \right]  \\
   & = & \tr \, \Phi(\rho)
  \left[  \log \Phi(\rho) - \log \Phi(\gamma) \right] \\
   & = &  H\big[\Phi(\rho), \Phi(\gamma) \big] .
\end{eqnarray*}
It is tempting to follow our previous strategy and choose
$A = \rho$,  $\log B = \log \gamma + \widehat{\Phi}
   \big[ \log \Phi(\rho) - \log \Phi(\gamma) \big]$.
However, we have been unable to verify that
$ \tr e^ {\log \gamma +
\widehat{\Phi}  [ \log \Phi(\rho) - \log \Phi(\gamma) ]} \leq 1$
as required by this approach.

Instead, we use the representation (\ref{eq:Lind.rep}) or (\ref{eq:Vrep}).
Rather than applying the equality conditions in
Theorem~\ref{thm:mon.tr} directly to (\ref{eq:mono.via.pt}),
it is useful to repeat the argument  for an appropriate
choice of $A$ and $B$.

\pf Choose $A =   V \rho V^{\dg} $,
$\log B =  \log (V \gamma V^{\dg})
  +  \log \tr_2 \big( V \rho V^{\dg} \big)
  - \log \tr_2 \big(  V \gamma V^{\dg}    \big)$ where $V$
is again the partial isometry as in
(\ref{eq:Vrep}) of Section~\ref{sect:Lind.rep}.
$B$ is defined so that the last two terms in
$\log B$ are extended from ${\cal H}$ to ${\cal H} \ot {\bf C}^m$
so that $\ker(B) \subset \ker(A)$.
The condition for equality in (\ref{eq:mono.via.pt}) is then
\begin{eqnarray}  \label{eq:eqal.mono.form.a}
 \log ( V \rho V^{\dg}  )
    -  \log (V \gamma V^{\dg} )   & = &
 \log  \tr_2  (V \rho V^{\dg} )
  - \log \tr_2 (V \gamma V^{\dg} )
 \\
  & = &  \log \Phi(\rho) - \log \Phi(\gamma)   \nonumber
\end{eqnarray}
We can put this into a more useful form by noting that for
 a partial isometry $V$
\begin{eqnarray}
 \log \, ( V \rho V^{\dg}  )  -  \log \, (V \gamma V^{\dg} ) =
    V \Big[ \log \rho - \log \gamma  \Big]  V^{\dg}
\end{eqnarray}
from which it follows that (\ref{eq:eqal.mono.form.a}) is equivalent to
\begin{eqnarray} \label{eq:cond.a}
  V \Big[ \log \rho - \log \gamma  \Big]  V^{\dg}
 = \log \Phi(\rho) - \log \Phi(\gamma)   .
\end{eqnarray}
Multiplying by $V^{\dg} $ on the left and $V$ on the right
and using that $V^{\dg}V = I$, one sees that (\ref{eq:cond.a}) implies
\begin{eqnarray} \label{eq:cond.b}
 \log \rho - \log \gamma
 = V^{\dg} \Big[ \log \Phi(\rho) - \log \Phi(\gamma)  \Big]  V .
\end{eqnarray}
Taking the partial trace $\tr_2$ over the auxiliary space in
 (\ref{eq:cond.b})  yields  (\ref{eq:eqal.gen.mono})
since $\wh{\Phi}(P) = \sum_k F_k^{\dg} P F_k = V^{\dg} P V$
for all $P$ in ${\cal H}$.   ~~~\qed

Another useful necessary condition for equality in
(\ref{eq:mono}) can be obtained by multiplying
both sides of (\ref{eq:cond.a}) by the projection
$ VV^{\dg}$.  Since $ V^{\dg}V = I$, one finds
\begin{eqnarray}  \label{eq:imp.proj}
 VV^{\dg} \Big[ \log \Phi(\rho) -  \log \Phi(\gamma) \Big]  & = &
V \Big[ \log \rho - \log \gamma  \Big]  V^{\dg}  \nonumber \\
 & = &  \Big[ \log \Phi(\rho) -  \log \Phi(\gamma) \Big] VV^{\dg}
\end{eqnarray}
i.e., the projection $VV^{\dg}$ commutes with
$\big[ \log \Phi(\rho) -  \log \Phi(\gamma) \big] $.
Taking the partial trace and noting that
$\Phi(I) = \tr_2 VV^{\dg} $
we can summarize this discussion in the following
\begin{cor} \label{cor:eqal.mon.nec2}
If equality holds in {\em (\ref{eq:mono})}, then
\begin{eqnarray}   \label{eq:cond.c}
 \Phi \big( \log \rho - \log \gamma \big) =
  \Phi(I) \left[  \log \Phi(\rho) - \log \Phi(\gamma) \right]
  = \left[  \log \Phi(\rho) - \log \Phi(\gamma) \right]  \Phi(I).
\end{eqnarray}
Moreover,
$   \log \Phi(\rho) - \log \Phi(\gamma) $
commutes with the projection
$VV^{\dg} = \sum_{k, \ell} |k \kb \ell |  F_k F_{\ell}^{\dg}$ where
$\{ F_k\}$ is a set of Kraus operators for $\Phi$, i.e.,
$\Phi(\rho) = \sum_k F_k \rho F_k^{\dg}$ and $| k \ket$ is an
orthonormal basis for the auxiliary space ${\cal H}_2$.
\end{cor}

The results of this section also hold in the more general
situation when
$\Phi : {\cal B}({\cal H}_A) \mapsto {\cal B}({\cal H}_A^{\prime})$
maps operators on one Hilbert space to those on another, in
which case $F_k : {\cal H}_A) \mapsto {\cal H}_A^{\prime}$.


\section{The Holevo bound}  \label{sect:Holv}

\subsection{Background}

One reason for studying conditions for equality is that other results,
 such as Holevo's celebrated bound~\cite{Hv0} on the accessible
information, can be obtained rather easily
from SSA or some form of the monotonicity of relative
entropy.  However, obtaining the
corresponding conditions for equality is not as
straightforward as one might hope
because of the need to introduce an auxiliary system.
Although Holevo's  bound is quite general, it is often applied
in situations where  $\wt{\rho}_j = \Phi(\rho_j)$ is the output
of a noisy quantum channel $\Phi$ with input $\rho_j$.  We
use the tilde $\wt{}$  as a reminder of this, as well as to
ensure a distinction from other density matrices which arise.

  For any fixed POVM and density
matrix $\gamma$, $p(b) = \tr \, (\gamma E_b)$ defines a classical
probability distribution whose entropy we denote
$S[\tr \, \gamma E_b]$.  The Holevo bound states that for
any ensemble of density matrices ${\cal E} = \{ \pi_j \wt{\rho}_j \}$
with average density matrix $\wt{\rho} = \sum_j \pi_j \wt{\rho}_j$
the accessible information in the ensemble satisfies
\begin{eqnarray}  \label{eq:acc.inf}
   I(\cale, \calm) & \equiv & 
  S[\tr \, \wt{\rho} E_b] -  \sum_j \pi_j  \, S[\tr \, \wt{\rho}_j E_b]
 \\ & \leq & S(\wt{\rho}) - \sum_j \pi_j S(\wt{\rho}_j) \label{eq:Holv}
\end{eqnarray}
for any POVM $\calm = \{ E_b \}$..  If all of the $\wt{\rho}_j$ 
commute, then it is easy to see that equality can be achieved by choosing
the $E_b$ to be the spectral projections which simultaneously diagonalize
the density matrices $\wt{\rho}_j$.  We wish to show that this
condition is also necessary, i.e., equality can only be achieved in
(\ref{eq:Holv}) if  all the $\wt{\rho}_j$ commute.

It is known~\cite{KR2,YO} that
(\ref{eq:Holv}) can be obtained  from~(\ref{eq:mono}).
First, observe that
\begin{eqnarray} \label{eq:Holv.equiv}
  S(\wt{\rho}) - \sum_j \pi_j S(\wt{\rho}_j) =
\sum_j \pi_j H(\wt{\rho}_j, \wt{\rho})
\end{eqnarray}
Now let $  \Omega_{\calm}$ be the map
  $\Omega_{\calm}(A)
=  \sum_b |b \kb b| \, \tr (A E_b)$ where $\calm = \{ E_b \}$.
 Then $\Omega_{\calm}$ is a stochastic map of the
special type known as a Q-C channel and
the Holevo bound
(\ref{eq:Holv}) follows immediately from (\ref{eq:Holv.equiv}) and
\begin{eqnarray}  \label{eq:Holv.Omega}
H[\Omega_{\calm}(\wt{\rho}_j), \Omega_{\calm}(\wt{\rho}) ]
\leq H(\wt{\rho}_j,\wt{\rho}).
\end{eqnarray}

\subsection{Equality conditions}  \label{sect:eqal.Holv}
We will henceforth assume that $\{ \pi_j ,\wt{\rho}_j\}$ is a
fixed ensemble and seek conditions under which we can find a POVM
satisfying the equality requirements.
Since $\wh{\Omega}_{\calm} (D) =  \sum_b E_b \bra b , D b \ket$,
 applying Theorem~\ref{thm:eqal.gen.mono}  yields
 conditions for equality in
(\ref{eq:Holv.Omega}).  For equality in
(\ref{eq:Holv})  these conditions must hold for every $j$ and reduce to
\begin{eqnarray}  \label{eq:cond.g}
    \log\wt{\rho}_j- \log \wt{\rho}  = \sum_b \, E_b ~
   \log \frac{\tr E_b \wt{\rho}_j }{\tr E_b \wt{\rho}}
    ~~ \forall ~ j
\end{eqnarray}
where this should be interpreted as a condition on
$\ker(\wt{\rho}_j)^\perp$ in which case all terms
are well-defined.  (Indeed, since the condition arises from the
use of Klein's inequality and the requirement $A = B$, the
operators in $B$ must be defined to be zero on $\ker(A)$,
which reduces to $\ker(\wt{\rho}_j)$ in the situation considered
here.)  If the POVM $\{ E_b \}$
consists of a set of mutually orthogonal projections, then it is
immediate that the operators
$Z_j \equiv  \log\wt{\rho}_j- \log \wt{\rho} $ commute, since
(\ref{eq:cond.g}) can be regarded as the spectral decomposition
of $Z_j$.  To show that the $\wt{\rho}_j$ themselves commute,
observe that
\begin{eqnarray*}
  1 = \tr \, \wt{\rho}_j & = &
\tr \, e^{log \wt{\rho}  + \, [\log \wt{\rho}_j - \log \wt{\rho}]} \\
& \leq & \tr \, \wt{\rho} ~ e^{\log \wt{\rho}_j - \log \wt{\rho}} \\
    & = &   \tr \, \wt{\rho} ~ e^{\sum_b \, E_b
   \log \frac{\tr E_b \wt{\rho}_j }{\tr E_b \wt{\rho} } } \\
   & = & \tr \, \wt{\rho} \,  \sum_b \, E_b
    \frac{\tr E_b \wt{\rho}_j }{\tr E_b \wt{\rho} } \\
    & = &  \sum_b \tr E_b \wt{\rho}_j = 1
\end{eqnarray*}
where we have used Theorem~\ref{thm:gold} with
$A = \log \wt{\rho}, B = \log \wt{\rho}_j - \log \wt{\rho}$,
 and the fact that for orthogonal projections
$e^{\sum_b a_b E_b} = \sum_b e^{a_b} E_b$.  The conditions
for equality in Theorem~\ref{thm:gold} then imply that
$\log \wt{\rho}_j$ and $\log \wt{\rho}$ commute for all $j$.
Hence  $\wt{\rho}_j$ and $\wt{\rho}_k$ also commute for all
$j,k$ when the POVM consists of mutually orthogonal projections.

Using King's observation in the next section, one can reduce
the general case to that of projective measurements.  However,
we prefer to use the equality conditions to show directly that
the elements of the POVM must be orthogonal.
Moreover, the commutativity condition
involving $VV^{\dg}$ is reminiscent of the more
sophisticated Connes cocyle approach used by Petz, and thus
of some interest.

  Since the Kraus operators
for the Q-C map $\Omega_{\calm}$ can be chosen  as
$F_{kb} = |b \kb k | \sqrt{E_b}$ where $|b \ket$ and $|k \ket$
are orthonormal bases, one finds
\begin{eqnarray}  \label{eq:proj.cond}
   VV^{\dg}  = \sum_{b,c} \sum_{k, \ell}
     | b \kb c|  \, \bra k \sqrt{E_b} \sqrt{E_c} \, \ell \ket
  =  \sum_{b,c} | b \kb c| \, \bra \phi \sqrt{E_b} \sqrt{E_c} \, \phi \ket
.
\end{eqnarray}
By (\ref{eq:imp.proj}), this must commute for all $j$ with $\log
\Omega_{\calm}(\wt{\rho}_j) -
   \log \Omega_{\calm}(\wt{\rho}_j)  $ which can be
written in the form $ \sum_b z_{bj}|b \kb b| $
with  $z_{bj} =  \log \frac{\tr E_b\wt{\rho}_j }{\tr E_b \wt{\rho}} $.
A diagonal operator of the form $ \sum_b  z_b|b \kb b|$ with
all $z_b \neq 0$ will commute with the projection in
(\ref{eq:proj.cond}) if and only if all off-diagonal terms
are zero.  This will hold if the POVM is a projective
measurement, since then
$\sqrt{E_b} \sqrt{E_c} = E_b E_c = E_b \delta_{bc}$.
To see that this is necessary, note that the possibility that
the vector $\phi$ is orthogonal to all $E_b$ is precluded
by the condition that $\sum_b E_b = I$.  Moreover, since the
orthonormal basis $|k \ket$ is arbitrary, $\phi$ can be chosen
to be arbitrary.  The restriction that (\ref{eq:cond.g}) hold
only on $\ker(\wt{\rho}_j)^\perp$ may permit
some $z_{bj} = 0$; however, for each $b$ there will always
be at least one $j$ for which $z_{bj} \neq 0$, and this suffices.
~~~\qed

\medskip

One can obtain an alternate form of the equality conditions
from  Corollary~\ref{cor:eqal.mon.nec2}.
Since $\Phi(I) = \sum_b |b \kb b|  \tr E_b $,
another necessary condition for equality in
(\ref{eq:Holv}) is
\begin{eqnarray} \label{eq:cond.e}
  \tr E_b \left[ \log \wt{\rho}_j  - \log \wt{\rho} \right] =
  \tr E_b  \Big(  \log \tr E_b \wt{\rho}_j -
       \log \tr E_b \wt{\rho} \Big)~~~~\forall ~j, ~b
\end{eqnarray}
 Inserting this in (\ref{eq:cond.g})
yields the requirement
\begin{eqnarray}  \label{eq:cond.f}
\log\wt{\rho}_j- \log \wt{\rho}  & = &  \sum_b  \frac{1}{\tr E_b}
   E_b  \tr \, E_b \, \big[ \log\wt{\rho}_j- \log \wt{\rho} \big]
\end{eqnarray}
which can be rewritten as
\begin{eqnarray} \label{eq:cond.Z}
Z_j =  \sum_b \frac{ | E_b \ket }{\tr E_b} \, \bra E_b,  Z_j \ket
~~~\forall j
\end{eqnarray}  where  $Z_j =  \log\wt{\rho}_j- \log \wt{\rho} $ and
the bra-ket now refer to the Hilbert-Schmidt inner product.
This implies that $ \sum_b \frac{ | E_b \kb E_b | }{\tr E_b}$
projects onto the span($\{ Z_j \}$).   However, this alone
is not sufficient to
imply that the $E_b$ form a projective measurement.

\subsection{Other approaches}

 Chris King has observed \cite{King3} that when
the POVM is a projective measurement of the form $E_b = |b \kb b|$,
one can obtain the Holevo bound from the joint convexity of
relative entropy.   Let
$\beta(\wt{\rho}) = \sum_b |b \kb b| \tr E_b \wt{\rho}$.
Then applying Theorem~\ref{thm:relent.conv} to
$H[\wt{\rho}, \beta(\wt{\rho})]$
yields
\begin{eqnarray}  \label{eq:Holv.JC}
 & ~ & -S(\wt{\rho}) + S(\tr E_b \wt{\rho}) \leq  \sum_j \pi_j
     \Big[  - S(\wt{\rho}_j) + S(\tr E_b \wt{\rho}_j) \Big] \\
{\rm or} & ~ &   \nonumber \\
& ~ & S(\tr E_b \wt{\rho}) -  \sum_j \pi_j S(\tr E_b \wt{\rho}_j) \leq
        S(\wt{\rho}) -  \sum_j \pi_jS(\wt{\rho}_j)  \nonumber
\end{eqnarray}
with equality if and only if
\begin{eqnarray} \label{eq:JC.Hv.cond.a}
  \log \wt{\rho} - \sum_b |b \kb b| \log \tr E_b \wt{\rho} =
     \log \wt{\rho}_j - \sum_b |b \kb b| \log \tr E_b \wt{\rho}_j
~~~\forall ~ j.
\end{eqnarray}
This is equivalent to (\ref{eq:cond.g}) when  $E_b = |b \kb b|$,
and the argument can be extended to more general projective
measurements.

King also pointed out that if $\{ E_b \}$ is an arbitrary POVM, the
construction in Section~\ref{sect:POVM} can be used to show that
(\ref{eq:Holv}) and (\ref{eq:cond.g}) are equivalent to the equalities
obtained when $\wt{\rho}_j$ is replaced by $V \wt{\rho}_j V^{\dg}$ and
$E_b$ by $F_b$.  Since the $\{ F_b \}$ form a projective
measurement, we can conclude from the argument above that
equality implies that all $V\wt{\rho}_jV^{\dg}$ commute,
which implies that all $\wt{\rho}_j$ also commute since
$V^{\dg} V = I$.

It should be noted that Petz was able to use his equality conditions
to find the conditions for equality in the Holevo bound
and this is sketched in ~\cite{OPW}.  Indeed,  Petz's
analogue of (\ref{eq:JC.Hv.cond.a}) is
$\wt{\rho}^{it} D^{-it} = \wt{\rho}_j^{it} D_j^{-it} ~~ \forall ~j$
where $D, D_j$ denotes the diagonal parts of $\wt{\rho}, \wt{\rho}_j$
respectively.    Then
\begin{eqnarray}  \label{eq:Petz.Hv}
\wt{\rho}_j^{it} = \wt{\rho}^{it} D^{-it} D_j^{it} .
\end{eqnarray}
Since (\ref{eq:Petz.Hv})  holds for all real $t$, as well as all $j$,
it also implies
$\wt{\rho}_j^{\, -it} = \wt{\rho}^{\,-it} D^{it} D_j^{-it}$.;
However, taking the adjoint of (\ref{eq:Petz.Hv})
yields $\wt{\rho}_j^{\,-it} = D_j^{-it} D^{it} \wt{\rho}^{\,-it}  $.
Therefore, $\wt{\rho}^{\, -it}$ commutes with the diagonal matrix
$D^{it} D_j^{-it} = D_j^{-it} D^{it}$ and must also be diagonal.
This gives a simultaneous diagonalization of
all $\wt{\rho}_j^{it} $ which means that all $\wt{\rho}_j$ commute.

Holevo's original longer derivation \cite{Hv0} of the bound
(\ref{eq:Holv})  also concluded
that commutativity was necessary and sufficient for
equality.   Some simplifications of this argument
were given by Fuchs~\cite{Fuchs}  in his thesis.

\subsection{Another bound on accessible information} \label{sect:alt.bnd}
When $\rho$ is a density matrix, the mapping 
$A \mapsto \rho^{-1/2} A \rho^{-1/2}$ and its inverse gives
a duality between ensembles and POVM's. Hall~\cite{Hall}  observed
that this duality can be used to give
 another upper bound on the accessible
information (\ref{eq:acc.inf}) in terms of the POVM and average density
$\rho$, i.e.,
\begin{eqnarray}  \label{eq:alt.bnd}
   I(\cale, \calm)  &  \leq  & S(\rho) - \sum_b \tau_b  \,
  S\left( \ts{\frac{1}{\tau_b}}  \sqrt{\rho} \, E_b 
\sqrt{\rho} \right) \\
   & = & \sum_b \tau_b \, H\left(  \ts{\frac{1}{\tau_b}}  
  \sqrt{\rho} \, E_b  \sqrt{\rho} \, , \, \rho \right)
\end{eqnarray}
where $\tau_b = \tr\, E_b \rho$.  This inequality can be obtained  
from the monoticity of
relative entropy under the Q-C map $\Omega_{\cale}(A) = 
   \sum_j |j \kb j| \pi_j \rho^{-1/2} \rho_j \rho^{-1/2}$
 applied to $H\left(  \ts{\frac{1}{\tau_b}}  
  \sqrt{\rho} \, E_b  \sqrt{\rho} \, , \, \rho \right)$ as in
(\ref{eq:Holv.Omega}); or as in \cite{KR2} where an equivalent bound
was given.
The argument in Section~\ref{sect:eqal.Holv} can then be
used to show that equality can be
achieved  in (\ref{eq:alt.bnd}) if and
only if all $ \sqrt{\rho} \, E_b  \sqrt{\rho} $ commute.
Hall~\cite{Hall} also found this condition and noted that it
implies that $\rho$ commutes with every $E_b$ in the POVM.
 
One is often interested in (\ref{eq:Holv}) and
(\ref{eq:alt.bnd}) when
one wants to optimize the accessible information after using a noisy
quantum channel, $\Phi$.  It was observed in \cite{KR2} that, since
$\tr \Phi(\rho_j) E_b = \tr \rho_j \wh{\Phi}(E_b)$,
one can  regard the noise as either acting to transform
pure inputs $\rho_j$ to mixed state outputs $\Phi(\rho_j)$ {\em or}
as acting through the adjoint $\wh{\Phi}$ on the POVM with
uncorrupted outputs.   In the first case, one can bound the
right side of (\ref{eq:alt.bnd}) by choosing
the $E_b$ to be the spectral projections of the average
output state $\Phi(\rho)$ to yield
$I[\Phi(\cale), \calm] \leq S[\Phi(\rho)]$ which is weaker
than the corresponding Holevo bound.  Moreover, since the
optimal choice for
$\Phi(\rho_j)$ need not be in the image of $\Phi$,
it not necessarily
achievable even though the commutativity condition holds. 
Hall~\cite{Hall} discussed other situations in which the
bound can not be achieved despite the fact that 
all $ \sqrt{\rho} \, E_b  \sqrt{\rho} $ commute.

Viewing the noise as acting on the POVM, King and
Ruskai~\cite{KR2} defined 
 \begin{eqnarray}  \label{eq:UEP}
    U_{EP}(\Phi) = \sup_{\rho,\calm} \left[ S(\rho) - \sum_b \tau_b  \,
  S\left(  \ts{\frac{1}{\tau_b}} 
    \sqrt{\rho} \, \wh{\Phi}(E_b)   \sqrt{\rho} \right) \right]
\end{eqnarray}
with $\tau_b = \tr  \rho \wh{\Phi}(E_b) = \tr \Phi(\rho) E_b$.
If the supremum in (\ref{eq:UEP}) is achieved with an average
density and POVM for which $\sqrt{\rho} \, \wh{\Phi}(E_b) \sqrt{\rho}$
do not commute, then $U_{EP}(\Phi)$ is strictly greater than
the accessible information.   The questions of whether or not
(\ref{eq:UEP}) can actually exceed the optimal accessible
information, and how it might then be interpreted are 
under investigation.

 \pagebreak

\section{Concluding remarks}
\label{sect:conc}

The proof presented here for each inequality, SSA,
Theorem~\ref{thm:mon.tr}, Theorem~\ref{thm:relent.conv} and
the general monotonicity (\ref{eq:mono}), is quite short ---
only half a page using results from Section \ref{sec:tools} which
require less than one additional page {\em and} Theorem~\ref{exp.conc} .
However, as shown in the Appendix, even this result does
{\em not} require a long argument if one is permitted to use some
powerful tools of complex analysis.

It is certainly not unusual to find that complex analysis can
extremely be useful,  even when the functions of interest are real-valued.
Indeed, Lieb's original proof of the concavity of WYD  entropy
used a complex interpolation argument.  In his influential
book ~\cite{Sim} on Trace Ideals, Simon (extracting ideas from
Uhlmann~\cite{Uhl2})  gave a
longer``elementary'' proof using the Schwarz inequality,
perhaps inadvertently reinforcing the notion that any complete proof of
SSA is long and forbidding.  Similar ideas are implicit
in Ando~\cite{Ando} who restates the result in terms of
tensor product spaces and block matrices.  Uhlmann~\cite{Uhl2} again
demonstrated the power of complex interpolation by using it to prove
the monotonicity of relative entropy under completely positive
trace-preserving maps.  SSA then follows immediately as a special
case.  However, Uhlmann's approach, which has been extended
by Petz~\cite{Pz.mono,OP}, was developed within  the  framework
of the relative modular operator formalism developed by
Araki~\cite{Ak,BR,OP} for much more general situations.
Recently, Lesniewski and Ruskai~\cite{LesRu} observed that
within this relative modular operator framework,
monotonicity can be established directly using an argument
based on the Schwarz inequality.

The approach of this review is similar to that of Wehrl~\cite{W}
in that we view Theorem~\ref{exp.conc} as the ``essential ingredient''.
Indeed, Uhlmann~\cite{Uhl1,W}, using a
completely different approach, had independently recognized that
Theorem~\ref{exp.conc} would imply SSA.
However, Wehrl's otherwise excellent review stated  (at the
end of section III.B) that ``Unfortunately, the proof of [this] is not
easy at all.''  Later (in section III.C) Werhl again states that
``... the proof is surprisingly complicated.   I want to indicate
only that the concavity of $\tr \, e^{K + \log A}$ can be obtained
from Lieb's  theorem [on concavity of the WYD  entropy] through a
sequence of lemmas.''
Although aware that Epstein's approach~\cite{Ep}, which was
developed shortly after Lieb announced his results,
permitted a ``direct'' proof of Theorem~\ref{exp.conc}, Wehrl does not
seem to have fully appreciated it.   The utility of Epstein's
technique may  have been underestimated, in part,
because he presented his results in a form which applied to the full
collection of convex trace functions studied in~\cite{LbWYD}.  Checking
Epstein's hypotheses for the WYD function requires some
non-trivial mapping theorems.  This may have obscured the elegance of the
argument in Appendix~A.

It is worth noting that if the concavity of WYD  entropy
is regarded as the
key result, it is not necessary to use the long sequence of lemmas
Wehrl refers to in order prove SSA.   Lindblad~\cite{Lind74}
gave a direct proof of the joint convexity, Theorem~\ref{thm:relent.conv},
directly
by differentiating the WYD function.   Once this is done,
SSA follows via the purification argument sketched after
equation (\ref{eq:SSA.alt}) or, alternatively, the variant of
Uhlmann's argument described in \cite{Sim,W}.  Combining this with Lieb's
original complex interpolation proof of the concavity of the
WYD function, yields another ``short''
proof of SSA, albeit one which does not appear to be well-suited to
establishing conditions for equality.

Finally, we mention that Carlen and Lieb \cite{CarLb} obtained
another proof of SSA by using Epstein's technique to prove some
Minkowski type inequalities for $L_p$ trace norms.
Using a different approach, King \cite{King1,King2} recently proved
several additivity results for the minimal entropy and Holevo capacity
of a noisy channel by using $L_p$ inequalities in which
Epstein's technique provided a critical estimate.
This suggests that connections with $L_p$ inequalities, as
advocated by  Amosov, Holevo and Werner~\cite{AHW}, may be
a promising avenue for studying  entropy and
capacity in quantum information.
Despite the results mentioned above, many open conjectures remain;
see \cite{AHW,CarLb,King1,King2,WH} for further details.

\bigskip


\section*{Acknowledgments}

The work of M.B. Ruskai was partially supported  by
 the National Security Agency (NSA) and
 Advanced Research and Development Activity (ARDA) under
Army Research Office (ARO) contract number
   DAAG55-98-1-0374 and by the National Science
        Foundation under Grant numbers DMS-9706981 and DMS-0074566.
This paper is the result of questions first posed to the author
by M. Nielsen and it is a  pleasure to acknowledge the value
of  extensive discussions with him.
Some of these discussions took place, and parts of this paper
were written, when MBR and MN were visiting the
Institute for Theoretical Physics at the University of
California, Santa Barbara and thereby also partly supported by
the National Science Foundation under Grant PHY-9907949.

The author is extremely grateful to C. King for careful readings of
earlier versions of the manuscript leading to many useful comments,
and for communicating (and allowing to be included)
several critical observations needed
to obtain the equality conditions for the Holevo bound in
Section~\ref{sect:Holv}.  It is also a pleasure to thank
C. Fuchs
for explaining the history of the equality conditions in the
Holevo bound, E. Lieb to clarifying remarks about Bauman's
responsibility for the actual conjecture of the concavity of the WYD
function, D. Petz for discussions about his approach, and 
P. Shor for asking a question which stimulated 
the analysis  in Section~\ref{sect:alt.bnd}, and to M.J.W. Hall
for bringing Reference~\cite{Hall} to my attention.

\medskip

  \pagebreak

\appendix

\section{Epstein's proof of concavity of $A \raw \tr e^{K + \log A}$}

Let $f(x) = \tr e^{K + \log (A + xB)}$ with $A > 0$ strictly
positive and $K, B$ self-adjoint.  For sufficiently small
$x$, the function $f(x)$ is well-defined and the concavity
of  $F(A)$ in Theorem \ref{exp.conc} follows immediately if
 $f^{\prime\prime}(0) < 0$ for all choices of $B = B^*$.

Instead of dealing with $f$ directly, Epstein considered
the function $g(x) = x f(x^{-1})$ which is well-defined for
$|x| > \mu^{-1} \equiv \| A^{-1}\| \, \| B \| $ and can be
analytically continued to the upper half plane so that
\begin{eqnarray}
  g(z) =  \tr \, e^{K + \log (zA + B)}.
\end{eqnarray}
There are a number of equivalent (when meaningful) ways of
defining functions of matrices.   For the purposes needed
here it is natural to assume that the spectrum $\sigma(A)$
of the operator $A$ is contained in the domain of an
analytic function $F(z)$ and that
\begin{eqnarray}
  F(A) =  \frac{1}{2 \pi i} \oint \frac{F(z)}{zI - A} dz.
\end{eqnarray}

One can then  use the spectral mapping theorem
$\sigma [F(A)] \subset F[\sigma(A)]$ for an appropriate
sequence of functions to verify that
\begin{eqnarray*}
  \Im \, z > 0 & \Raw &  \Im \, \omega(zA + B) > 0 \\
  & \Raw &    \pi > \Im \, \omega[\log(zA + B)] > 0  \\
    & \Raw &    \pi > \Im \, \omega[K + \log(zA + B)] > 0  \\
  & \Raw &  \Im \,   \omega \left[e^{K + \log (zA + B)} \right] > 0  \\
 & \Raw &  \Im \,  \tr  \, e^{K + \log (zA + B)} > 0
\end{eqnarray*}
where $\Im $ denotes the imaginary part of a complex number and
$\omega$ is used to denote an
arbitrary element of the spectrum of the
indicated operator.  Thus, $g(z)$ maps the upper half plane
into the upper half plane.  Functions with this property
have been studied extensively under various names, including,
 ``operator monotone'', ``Herglotz'' or ``Pick''.  (See, for
example,~\cite{Ando,D,OP}).  It then follows that
$g$ has an integral representation of the form
\begin{eqnarray}
  g(z) =  a + bz +
 \int_{-\mu}^{\mu} \frac{1}{t - z}  \, dm(t)
\end{eqnarray}
for some positive measure $\mu(t)$.
This yields (via the change of variables $s = t^{-1}$)
\begin{eqnarray}
  f(x) & = &  ax + b + \int_{-\mu}^{\mu} \frac{x^2}{tx - 1} \, dm(t)
\end{eqnarray}
Differentiation under the integral sign can then be used to
establish that $f^{\prime\prime}(0) < 0$ as desired by observing
$ \frac{x^2}{tx - 1} = t^{-2} [ (xt + 1) + (xt -1)^{-1} ].$ \qed

\bigskip



{~~}

\end{document}